\documentclass[preprint2]{aastex}

\newcommand{\bfm}[1]{\mbox{\boldmath $#1$}}
\newcommand{\mat}[1]{\mbox{\boldmath $\mathsf{#1}$}}
\newcommand{\pcm}{\,cm$^{-3}$}
\newcommand{\kms}{\,km\,s$^{-1}$}

\shorttitle{Turbulence in molecular clouds}
\shortauthors{Downes \& O'Sullivan}

\begin{document}


\title{Non-ideal MHD turbulent decay in molecular clouds}


\author{T.P. Downes\altaffilmark{1,2} and S. O'Sullivan\altaffilmark{2}}
\affil{School of Mathematical Sciences, Dublin City University,
	Glasnevin, Dublin 9, Ireland}
	\email{turlough.downes@dcu.ie}

\altaffiltext{1}{School of Cosmic Physics, Dublin Institute for Advanced
Studies, 31 Fitzwilliam Place, Dublin 2, Ireland}
\altaffiltext{2}{National Centre for Plasma Science and Technology,
	Dublin City University, Glasnevin, Dublin 9, Ireland}


\begin{abstract}
It is well known that non-ideal magnetohydrodynamic effects are
important in the dynamics of molecular clouds: both ambipolar
diffusion and possibly the Hall effect have been identified as significant.  
We present the results of a suite of simulations with a resolution of 
$512^3$ of turbulent decay in molecular clouds incorporating a simplified form 
of both ambipolar diffusion and the Hall effect simultaneously.  The initial
velocity field in the turbulence is varied from being super-Alfv\'enic and
hypersonic, through to trans-Alfv\'enic but still supersonic.

We find that ambipolar diffusion increases the rate of decay of
the turbulence increasing the decay from $t^{-1.25}$ to $t^{-1.4}$.  The Hall 
effect has virtually no impact in this regard.  The power spectra of density, 
velocity and the magnetic field are all affected by the non-ideal terms, 
being steepened significantly when compared with ideal MHD turbulence with 
exponents.  The density
power spectra components change from $\sim1.4$ to $\sim2.1$ for the ideal and
non-ideal simulations respectively, and power spectra of the other
variables all show similar modifications when non-ideal effects are
considered.  Again, the dominant source of these changes is ambipolar
diffusion rather than the Hall effect.  There is also a decoupling between 
the velocity field and the magnetic field at short length scales.  The Hall 
effect leads to enhanced magnetic reconnection, and hence less power, at 
short length scales.  The dependence of the velocity dispersion on the 
characteristic length scale is studied and found not to be power-law in nature.

\end{abstract}


\keywords{MHD — ISM: kinematics and dynamics — ISM: magnetic fields —
methods: numerical — turbulence}



\section{Introduction}
\label{sec:intro}

%
%

The role of turbulence in molecular cloud evolution has been a subject of
much study in the literature (see, for example, the excellent reviews of 
\citealt{mac04, elm04}).  Observations of the properties of
gas and dust motions in molecular clouds \citep{lar81} suggest that, 
indeed, turbulence is present.  It is clear that turbulent motion could
influence the star formation rate and efficiency as well as the initial
mass function \citep{elm93, kle03}.  Given all this there are several 
interesting questions which arise:
\begin{enumerate}
\item What is the source of molecular cloud turbulence?
\item How fast does it decay?
\item How does it affect star formation?
\item How does it affect the evolution of molecular clouds?
\end{enumerate}
The first two questions are clearly inter-related: if the turbulence
decays very quickly then we need a lot of energy from its source in
order to maintain it.  Indeed, to address the final two questions we
must first gain insight into the first two.

%
%

In order to study turbulence in molecular clouds we must resort to
numerical simulations.  Quite simply, there are no satisfactory analytic
techniques for addressing compressible magnetohydrodynamic (MHD)
turbulence \citep[e.g.][]{elm04}. 

Many authors have performed sophisticated numerical simulations in order to 
investigate both the qualitative nature of MHD turbulence and its decay
\citep{maclow98, maclow99, ost01, ves03, gus06, glover07, lem08, lem09}.  
Most of this work has been carried out for the case where ideal MHD is valid 
- i.e. on relatively large length scales.  When smaller length scales are 
considered (e.g. lengths of significantly less than a parsec) ambipolar 
diffusion becomes non-negligible in molecular clouds \citep{ois06}.  Some 
authors \citep{li08, kud08, ois06} have studied driven MHD turbulence in the 
presence of ambipolar diffusion.  All these authors find that ambipolar 
diffusion produces significant differences in the properties of the 
turbulence.

%
%

It has been suggested \citep{wardle04} that although the Hall resistivity is
generally at least an order of magnitude lower than the ambipolar
resistivity in molecular clouds, its effect should not be ignored due to the 
qualitative change it induces in the behavior of the magnetic field.  
Researchers working on reconnection and the solar wind have studied the Hall 
effect in the context of turbulence and found that, although the overall decay 
rate appears not to be affected, the usual coincidence of the magnetic and
velocity fields seen in MHD does not occur at small scales 
\citep{ser07, min06, mat03}.  Almost no work has been done on comparing
the influences of this effect coupled with that of ambipolar diffusion on 
turbulence (with the exception of low resolution simulations by 
\citealt{dos08}).  In particular, to our knowledge the work presented here 
represents the first systematic study of molecular cloud turbulence 
incorporating both the Hall effect and ambipolar diffusion simultaneously.

%
%

The main aim of this work is to examine in detail the differences between the 
decay of ideal MHD turbulence and that of more realistic non-ideal MHD 
turbulence with a full tensor resistivity incorporating the effects of 
ambipolar diffusion, the Hall effect and Ohmic resistivity.  This work
is new in two respects: no previous work has focussed on {\em decaying}
(i.e.\ un-driven) turbulence in the presence of non-ideal terms and, in 
addition, no previous work has addressed the issue of turbulence in the 
presence of both ambipolar diffusion and the Hall effect simultaneously.  This 
is the first of a short series of papers describing a comprehensive 
study of multifluid MHD turbulence in a parameter regime appropriate to 
molecular clouds.

%
%

In section \ref{sec:num-method} we outline the numerical techniques used
in this work, as well as the initial conditions and general set-up for the 
simulations while in section \ref{sec:analysis} we describe the methods
used to analyze the simulation data.  In section \ref{sec:results} we 
present and discuss the results of our simulations of turbulent decay.  
Finally, section \ref{sec:conclusions} contains a summary of our results.

\section{Numerical method}
\label{sec:num-method}
%
%

In this work we use the code HYDRA \citep{osd06, osd07} to integrate
the equations of non-ideal MHD (see section \ref{subsec:eqns-algorithm}).  We 
assume that the molecular cloud material we are simulating can be treated as 
isothermal and that initially the density and magnetic field are uniform.  For 
this work we assume spatially and temporally constant Ohmic, ambipolar and 
Hall resistivities (see section \ref{subsec:eqns-algorithm}).  

\subsection{Equations and algorithm}
\label{subsec:eqns-algorithm}
%
%

We briefly outline the equations and assumptions in our model here but
refer the reader to \cite{osd06, osd07} for a comprehensive description of both
the full abilities of the HYDRA code and the assumptions
underlying the equations used.

We assume that the cloud material can be treated as weakly
ionized.  This is clearly valid for molecular clouds and allows us to ignore 
the inertia of the charged species \citep{cio02, falle03}.  The equations 
solved in this work are then
\begin{eqnarray}
\frac{\partial \rho}{\partial t} + \nabla\cdot \left(\rho
		   \bfm{q}\right)  & = & 0 , \label{mass} \\
\frac{\partial \rho \bfm{q}}{\partial t}
	+ \nabla\cdot\left( \rho \bfm{q} \bfm{q} + a^2\rho\mat{I}\right) &
	= & \bfm{J}\times\bfm{B} , \label{neutral_mom} \\
\frac{\partial \bfm{B}}{\partial t} + 
\nabla\cdot(\bfm{q}\bfm{B}-\bfm{B}\bfm{q}) & = &
-\nabla\times\bfm{E}' \label{B_eqn} \\
\nabla\cdot\bfm{B} & = & 0 \label{divB} \\
\nabla\times\bfm{B} & = & \bfm{J} \label{eqn-J}
\end{eqnarray}
where $\rho$, $\bfm{q}$, $a$, $\mat{I}$, $\bfm{B}$ and $\bfm{J}$ are the neutral
mass density, neutral velocity, sound speed, identity matrix, magnetic field 
and current density respectively.  The electric field in the frame of the fluid,
$\bfm{E}'$, is calculated from the generalized Ohm's law for weakly
ionized fluids (e.g. \citealt{falle03, osd06}) and is given by
\begin{equation}
\bfm{E}' = \bfm{E}_{\rm O} +\bfm{E}_{\rm H} +\bfm{E}_{\rm A} ,
	\label{eqn_E1}
\end{equation}
where
\begin{eqnarray}
\bfm{E}_{\rm O} & = & (\bfm{J}\cdot\bfm{a}_{\rm O})\bfm{a}_{\rm O} ,
	\label{eqn_EO1} \\
\bfm{E}_{\rm H} & = & \bfm{J}\times\bfm{a}_{\rm H} , \label{eqn_EH1} \\
\bfm{E}_{\rm A} & = & -(\bfm{J}\times\bfm{a}_{\rm A})\times\bfm{a}_{\rm A}
	, \label{eqn_EA1} 
\end{eqnarray}
using the definitions $\bfm{a}_{\rm O}\equiv f_{\rm O} \bfm{B}$,
$\bfm{a}_{\rm H}\equiv f_{\rm H} \bfm{B}$, $\bfm{a}_{\rm A}\equiv
f_{\rm A} \bfm{B}$, where $f_{\rm O}\equiv \sqrt{r_{\rm O}}/B$,
$f_{\rm H}\equiv r_{\rm H}/B$, $f_{\rm A}\equiv \sqrt{r_{\rm
A}}/B$. $r_{\rm O}$, $r_{\rm H}$ and $r_{\rm A}$ are the Ohmic, Hall and 
ambipolar resistivities respectively.

In this work these resistivities are kept constant in both space and time.  We 
note that, physically, they depend on both the magnetic field and the
density of the various charged species in the fluid \citep[e.g.][]{cio02,
falle03, osd06, osd07} and hence, in reality, do vary in both space and 
time.  Treating the resistivities in such a simple way allows us to
gain a deeper understanding of their influence on turbulence without
having to consider the complicating effects of dynamically varying
resistivities at the same time.  As such, and as a first step away from the 
approximation of ideal MHD and a single form of magnetic diffusion we believe 
this is an interesting study.  Having gained some insight into
this simplified model a follow-up paper will address multifluid MHD 
turbulence under the influence of self-consistently calculated resistivities.

As noted by \citet{falle03} and \citet{osd06}, the main difficulty with 
standard numerical techniques for integrating equation \ref{B_eqn} lies with 
the Hall term.  As this term becomes dominant the stable time-step goes to
zero.  However, O'Sullivan \& Downes (\citeyear{osd06, osd07}) presented a 
novel, explicit numerical method for integrating this term such that the limit 
on the stable time-step is not overly restrictive.  We use this 
``Hall Diffusion Scheme'' in this work.  Of course, all explicitly differenced 
diffusion terms give rise to a stable time-step which is proportional to 
$\Delta x^2$, where $\Delta x$ is the resolution of the simulation.  To 
ameliorate this we use standard sub-cycling of the Hall terms and super 
time-stepping to accelerate the ambipolar diffusion terms 
\citep[see][]{ale96, osd06, osd07}.

Equations \ref{mass} -- \ref{B_eqn} are solved using a standard
shock-capturing, second order, finite volume, conservative scheme.  
Equation \ref{divB} is enforced using the method of \citet{dedner02}.  The 
effects of the diffusive terms are then incorporated in an operator split 
fashion. 

\subsection{Initial conditions}
\label{subsec:initial-conditions}
%
%
We examine the decay of supersonic MHD turbulence in conditions suitable for 
dense regions of molecular clouds.  While the simulations presented here are, 
of course, scale-free we present the initial conditions used in standard 
astrophysical units for ease of reading.  

The simulations are carried out in a cube of side $L=0.2$\,pc with
periodic boundary conditions being enforced on all faces.  The sound
speed is set to 0.55\kms and the initial density is chosen to be uniform
with a value of $10^6$\pcm.  The magnetic field is also initially
uniform in the $(1,1,1)$ direction with a magnitude of 1\,mG.  For these
conditions, suitable conductivities are $\sigma_{\rm O} = 1\times10^{10}$
\,s$^{-1}$, $\sigma_{\rm H} = 10^{-2}$\,s$^{-1}$ and 
$\sigma_{\rm A} = 10^{-1}$\,s$^{-1}$ (see figure 1, \citealt{wardle99}).
We choose these particular physical conditions with a view to maximizing the 
influence of the Hall effect in our simulations \citep{wardle99}.  In
this way we can use our simulations to find whether the Hall effect is ever
likely to be important in molecular cloud turbulence.

The initial velocity field is used to instigate the turbulence in these
simulations.  Each component of the velocity field is defined to be the sum 
of waves with 16 wave-vectors, each with random amplitude and phase - i.e. 
\begin{equation}
q_i(x,y,z) = \sum_{l,m,n=1}^{4} A^{lmn}_i
\cos(k^l_i x + k^m_i y + k^n_i z + \phi^{lmn}_i)
\end{equation}
where $i=0,1,2$ defines the component ($x$, $y$ or $z$ respectively) of the
velocity, $A^{lmn}_i$ and $\phi^{lmn}_i$ are the random amplitudes and phases and
\begin{equation}
k^l_i \equiv \frac{2 \pi l }{L} (1-\delta_{li})
\end{equation}
where $\delta_{li}$ is the usual Dirac delta function.  The inclusion of
the term in parenthesis in the definition of $k^l_i$ restricts the initial 
velocity field to be solenoidal.  Note that we perform all the analysis of
these simulations at $t \geq 0.2\,t_{\rm c}$ (1 flow crossing time) at which 
time the effects of the precise initial conditions used should be negligible.

The nomenclature for the simulations is xx-ab-c where xx denotes the type of 
physics (e.g.\ a standard molecular cloud run is ``mc'', ideal MHD is 
``mhd'' etc), ab is the initial rms 
Mach number and c is the resolution used.  The initial root-mean-square (rms) of 
the field is chosen to be either Mach 2.5, 5 or 10 depending on the simulation in
question.  These correspond to Alfv\'enic Mach numbers of
approximately 0.96, 1.9 and 3.85 respectively.  In addition to the
non-ideal MHD simulations described we also run 4 further simulations.
The first is an ideal MHD simulation (mhd-5-512) which we use for comparison
purposes, another is a pure hydrodynamic simulation (hd-256-0.5),  and the 
other two (ambi-5-512 and hall-5-512) only incorporate one of ambipolar 
diffusion or the Hall effect, respectively.  We use these latter simulations 
to separate out the effects of each of these diffusions to better understand 
the physics occurring.  Table \ref{table:sim-defs} contains definitions of 
the various simulations used in this work.  

\begin{table}
\begin{center}
\caption{Definition of the initial conditions used in the simulations in
this work. \label{table:sim-defs}}
\begin{tabular}{lccl}
\tableline \tableline
Simulation & Mach number\tablenotemark{a} & Resolution & Comment \\
\tableline
mc-5-64 & 5 & $64^3$ & - \\
mc-5-128 & 5 & $128^3$ & - \\
mc-5-256 & 5 & $256^3$ & - \\
mc-5-512 & 5 & $512^3$ & - \\
mc-2.5-512 & 2.5 & $512^3$ & - \\
mc-10-512 & 10 & $512^3$ & - \\
ambi-5-512 & 5 & $512^3$ & $r_{\rm H}=0$ \\
hall-5-512 & 5 & $512^3$ & $r_{\rm A}=0$ \\
mhd-5-512 & 5 & $512^3$ & Ideal MHD \\
hd-5-256 & 5 & $256^3$ & Hydrodynamic \\
\tableline
\end{tabular}
\tablenotetext{a}{Initial rms Mach number of the flow}
\end{center}
\end{table}

\section{Analysis}
%
%
\label{sec:analysis}
In this section we discuss the method of analysis of the output of the
simulations described in section \ref{subsec:initial-conditions}.  The
main aim of this paper is to investigate the decay rate of supersonic
turbulence in molecular clouds.  Hence, the main analysis carried out on
the simulation results is the calculation of the volume-averaged kinetic, 
magnetic and total energy in the simulation as a function of time.  These 
quantities are defined as
\begin{mathletters}
\begin{eqnarray}
e_{\rm k} & = & < \rho |\bfm{q}|^2>
\label{kin-def}\\
e_{\rm b} & = & <\frac{|\bfm{B}|^2}{2}> - \frac{<\bfm{B}>^2}{2}
\label{b-def}\\
e_{\rm tot} & = & e_b + e_k \label{etot-def}
\end{eqnarray}
\end{mathletters}
where the angle brackets denote averaging over the computational domain.  Note 
that $e_{\rm b}$ is therefore the difference in the magnetic energy at the 
current time and the magnetic energy at $t=0$ (see, for example,
\citealt{ves03, lem09}), given the assumption that no external electromotive 
force is applied and that our boundary conditions are periodic.  We also 
calculate the mass-weighted average Mach number, defined by
\begin{equation}
M = \frac{1}{a}\left\{\sigma_x^2 + \sigma_y^2 + \sigma_z^2\right\}^{1/2}
\end{equation}
where $a$ is the sound speed and the mass-weighted velocity dispersions,
$\sigma_\alpha$, are defined by
\begin{equation}
\sigma_\alpha = \left\{\frac{\left<\rho
	q_\alpha^2\right>}{\left<\rho\right>}\right\}^{1/2}
\end{equation}
where $\alpha$ is either $x$, $y$ or $z$ and the angle brackets denote
averaging over the computational domain (see \citealt{lem09}).

In section \ref{sec:power-spectra} we present the power spectra for the 
velocity, density and magnetic field for each of the $512^3$ simulations.  
These spectra are calculated by taking the power spectrum in the $x$, $y$ and 
$z$ directions and then integrating the power for all $\bfm{k}$ satisfying
$k \leq |\bfm{k}| < k + dk$ for each $k$ with $dk=1$.  This gives us some 
insight into the scale of structures being formed by the turbulence for the 
various initial conditions and range of physics examined.

Finally, in section \ref{sec:vel-disp} we calculate the velocity
dispersion as a function of length scale, $l$.  For these purposes we 
define the velocity dispersion to be
\begin{equation}
\sigma(l) = \left\{<\sigma_x^2(l)>_{\rm domain} + <\sigma_y^2(l)>_{\rm
	domain} + <\sigma_z^2(l)>_{\rm domain}\right\}^{\frac{1}{2}}
\end{equation}
\noindent where
\begin{equation}
\sigma_\alpha(l) = \left\{\left<q_\alpha^2\right>_l-\left< q_\alpha
\right>_l^2\right\}^{\frac{1}{2}}
\end{equation}
where $<\cdot>_l$ indicates an average taken over a cube of side $l$ in
the simulation domain and $<\cdot>_{\rm domain}$ indicates averaging of the
quantity over all such non-overlapping cubes within the domain.

%
%

\section{Results}
\label{sec:results}

We now present the results of the simulations carried out.  Each
simulation was run for one sound crossing time, $t_{\rm c} =
3.56\times10^{4}$\,yrs, of the simulation domain.  

\begin{figure}
\epsscale{1.00}
\plotone{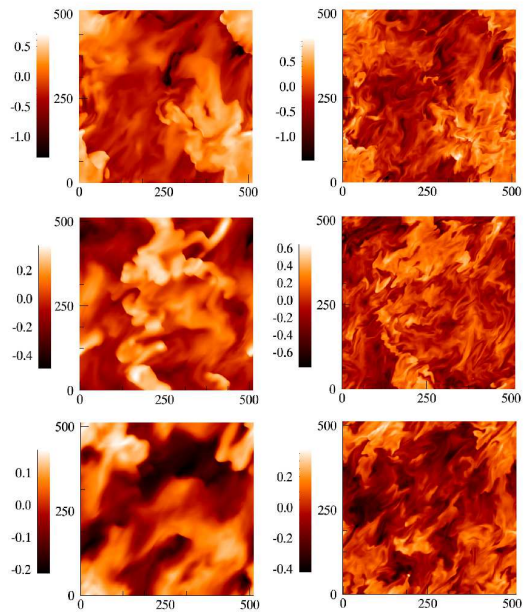}
\caption{Logscale plot of a slice in the mass density at times $t=0.2$,
$0.5$ and $1$\,$t_{\rm c}$ (top to bottom).  The left hand column is
the non-ideal simulation (mc-5-512) and the right hand column is the ideal 
MHD simulation (mhd-5-512). \label{fig:mc-mhd-den}}
\end{figure}

\begin{figure}
\epsscale{1.00}
\plotone{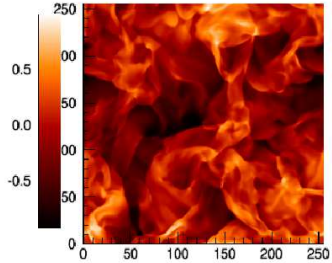}
\caption{Logscale plot of a slice in the mass density at time $t=0.2$\,
$t_{\rm c}$ for the hydrodynamic simulation (hd-5-256). \label{fig:hydro-den}}
\end{figure}

Figure \ref{fig:mc-mhd-den} shows the density distributions in a slice
at $x=0.1$\,pc (i.e. the mid-plane) and times $t=0.2$, $0.5$ and $1$\,
$t_{\rm c}$ for the mc-5-512 and the mhd-5-512 simulations.  Both the non-ideal 
and ideal MHD simulations show some anisotropy with respect to the
projected initial magnetic field direction - i.e. the $(1,1)$ direction - with 
filaments both perpendicular and parallel to this direction.  This is what 
would be expected as the material will flow preferentially along the $(1,1,1)$ 
direction and hence we expect the shock fronts to be normal to it.
This anisotropy becomes more evident as time progresses because
the kinetic energy decays to below the energy of the mean magnetic field
(which is conserved with time) - hence the flow evolves from being
dominated by kinetic energy to dominated by magnetic energy over the
lifetime of the simulations.  Figure \ref{fig:hydro-den} contains a plot of 
the same density distribution for hydro-5-256 at $t=0.2$\,$t_{\rm c}$ for 
comparison.  It can be seen that this is qualitatively different to both the 
ideal and non-ideal simulations shown in figure \ref{fig:mc-mhd-den} and does 
not display any signs of anisotropy.

There are obvious qualitative differences between the mc-5-512
and mhd-5-512 simulations at all times with the density distribution in the 
ideal case containing much more small scale structure than the non-ideal 
case.  This results from the ability of the magnetic field to move with 
respect to the fluid.  This means that it is not compressed as much as in the 
ideal case, and hence the same level of small structure is not formed.  Since
the coupling between the magnetic field and the flow field is strong, even
though it is imperfect, this lack of small-scale structure in the
magnetic field becomes mirrored in the density field.  The implications of this 
are discussed in more detail in section \ref{subsec:energy-decay}.
Clearly, in the non-ideal simulation energy is not as efficiently
transported from large scales to smaller scales as in the ideal case.

We now continue our discussion of the results by considering a resolution 
study to demonstrate convergence of our numerical results.  We then go on to 
discuss the properties of the density, velocity and magnetic field distribution resulting from the 
turbulence, and finally the properties of the energy decay.

\subsection{Resolution study}
%
%
\label{subsec:resolution-study}

In order to be reasonably confident that the results we present in this
work are converged we have performed a resolution study for the
conditions used in the simulations.  We have run simulations which are
identical in all respects except for the resolution, which varies from
$64^3$ up to $512^3$ (see table \ref{table:sim-defs}).  We then analyze
these in an identical fashion and consider the differences between our
results for the different resolutions.

\begin{figure}
\epsscale{0.80}
\plotone{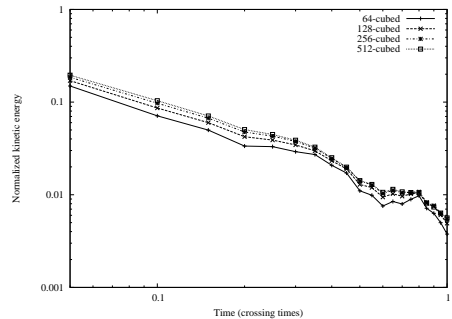}
\caption{Logscale plot of the kinetic energy (normalized to its initial
value) for each of the simulations in the resolution 
	study.\label{fig:res-energy}}
\end{figure}

Figure \ref{fig:res-energy} shows plots of the kinetic energy normalized
to its initial value as a function of time for each of the simulations in our 
resolution study.  The $64^3$ and $128^3$ plots do appear to be significantly 
different from the higher resolution simulations.  However, the $256^3$ and
$512^3$ simulations are much more similar with a maximum difference in
the total kinetic energy in the simulation at any one time being less
than 10\%.  Table \ref{table:decay-exp} contains least squares fits of
the decay over the time-interval $[0.2 t_{\rm c}, t_{\rm c}]$ assuming it to 
be of the form $t^{-\beta}$ for each of the
kinetic energy ($\beta_{\rm K}$), the magnetic energy ($\beta_{\rm B}$) and 
their total ($\beta_{\rm Tot}$).  It can be seen that $\beta_{\rm K}$ 
varies by approximately 4\% over the entire range of the resolution study 
(simulations mc-5-64 through mc-5-512).  We feel, therefore, that we can be 
fairly confident of the value of this exponent when comparing it with the other
simulations presented here.

\begin{table}
\begin{center}
\caption{The values of the exponent for the kinetic, magnetic and total 
energy decay for the simulations presented in this work.  These
exponents are calculated by fitting the data over the time interval
$[0.2 t_{\rm c}, t_{\rm c}]$.
	\label{table:decay-exp}}
\begin{tabular}{lccc}
\tableline \tableline
Simulation & $\beta_{\rm K}$ & $\beta_{\rm B}$ & $\beta_{\rm Tot}$ \\
\tableline
mc-5-64 & 1.34 & 1.40 & 1.35 \\
mc-5-128 & 1.33 & 1.35 & 1.34 \\ 
mc-5-256 & 1.37 & 1.36 & 1.36   \\ 
mc-5-512 & 1.40 & 1.37 & 1.39 \\ 
ambi-5-512 & 1.40 & 1.35 & 1.38 \\  
hall-5-512 & 1.25 & 1.18 & 1.22 \\ 
mhd-5-512 & 1.26 & 1.19 & 1.23 \\ 
hydro-5-256 & 1.10 & - & 1.10 \\ 
mc-2.5-512 & 1.21 & 1.29 & 1.23 \\
mc-10-512 & 1.42 & 1.39 & 1.41 \\ \tableline
\end{tabular}
\end{center}
\end{table}

The decay of the energy in the magnetic field is more rapid than that in
the kinetic energy with a difference between $\beta_{\rm K}$ and
$\beta_{\rm B}$ of about 4\%.  At low resolution (mc-5-64) 
$\beta_{\rm B}$ is at its highest which is a sign that the low resolution is 
introducing sufficient numerical viscosity to induce large amounts of numerical 
reconnection.  As the resolution is increased $\beta_{\rm B}$ reduces
to approximately 1.36 and stays at about this value even up to the
maximum resolution of $512^3$.  We can be reasonably confident then that
resolution is not affecting our estimate of the decay rate of the
magnetic energy at our maximum resolution.

The decay of the total energy, being derived from the decay of kinetic
and magnetic energy, is also reasonably well converged with a total
change over the entire range of the resolution study of around 4\%.

Note that, while we can be confident from our results that the decay
rate is converged, we have to be more careful when considering our power
spectra results presented in sections \ref{sec:power-spectra} and
\ref{sec:mach-effect-power-spectra}.  For example, \cite{krit07},
\cite{sch09} and \cite{lem09} present power spectra in ideal MHD which suggest 
that at resolutions of $1024^3$ the turbulent inertial range is established 
over at most a decade in $k$, while for resolutions of $512^3$ this falls to 
around half a decade.

\subsection{Energy decay}
\label{subsec:energy-decay}

We now discuss the behavior of the kinetic and magnetic energy in our
non-ideal simulations and compare them with those for our ideal
simulation.

\subsubsection{Kinetic energy decay}
\label{sec:kin-decay}

\begin{figure} 
\epsscale{0.80}
\plotone{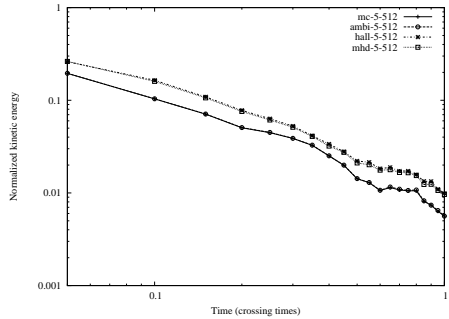}
\caption{Logscale plot of the kinetic energy (normalized to its initial
value) for simulations mc-5-512, ambi-5-512, hall-5-512 and mhd-5-512.
Note that the data for the mc-5-512 and ambi-5-512 are almost identical.
\label{fig:non-ideal-kin}}
\end{figure}

Figure \ref{fig:non-ideal-kin} contains plots of the kinetic energy as a
function of time for the mc-5-512, ambi-5-512, hall-5-512 and mhd-5-512
simulations.  It is apparent that the behavior of mhd-5-512 is rather similar 
to hall-5-512 and the behavior of ambi-5-512 is similar to that of
mc-5-512.  This indicates that, at least for the kinetic energy decay,
the Hall effect has little impact.  However, there is a marked
difference between those simulations incorporating ambipolar diffusion
(mc-5-512 and ambi-5-512) and those which do not (hall-5-512 and
mhd-5-512).  Ambipolar diffusion clearly increases the decay rate of the
turbulence.  This can also be seen from the data presented in table
\ref{table:decay-exp} where the exponents of the simulations containing
ambipolar diffusion are greater by about 10\% than those without.

In general we expect that turbulence will decay more rapidly in systems which 
have higher viscosity than those without.  Clearly the ambipolar diffusion,
although it corresponds to a ``viscosity'' in the magnetic field, 
acts in a similar way to the usual viscous forces in a fluid when
considering this system.  This can be seen in figure \ref{fig:mc-mhd-den} 
where the density features are more smeared out in the non-ideal case as would 
be expected if we simply introduced a viscous term into the momentum equations.

It is important to note here that the non-ideal diffusive terms in the 
induction equation will lead to enhanced reconnection in the magnetic field.  
Since we assume an isothermal equation of state there is no path by which the
energy released by reconnection can find its way to the kinetic energy
of the system.  There may be significant differences between the
effects of the non-ideal terms on the decay rate of turbulence in the 
isothermal and non-isothermal regimes in molecular clouds.

\subsubsection{Magnetic energy growth and decay}
\label{sec:b-decay}

\begin{figure} 
\epsscale{0.80}
\plotone{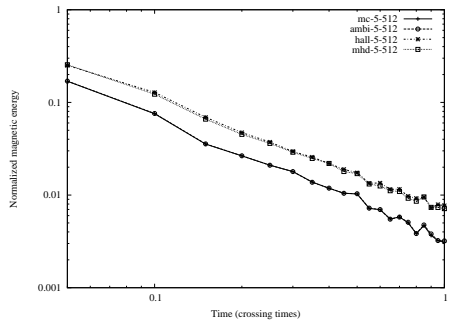}
\caption{Logscale plot of the magnetic energy (normalized to the initial
kinetic energy) for simulations mc-5-512, ambi-5-512, hall-5-512 and 
mhd-5-512.  Note that the data for mc-5-512 and ambi-5-512 are almost
identical.  See text.\label{fig:mhd-mc-b}}
\end{figure}

Figure \ref{fig:mhd-mc-b} contains plots of the energy of the magnetic
perturbations induced by the turbulence (normalized to the initial kinetic 
energy) as a function of time for the mc-5-512, ambi-5-512, hall-5-512 and 
mhd-5-512 simulations.  A similar trend is evident in the results for magnetic 
energy decay as that already noted in section \ref{sec:kin-decay}.  Again, the 
main result is that ambipolar diffusion increases the energy decay rate while
the Hall effect does little to influence it.  This result is borne out
by the data in table \ref{table:decay-exp} in which the decay exponents
are greater for the simulations incorporating ambipolar diffusion by
about 14\% than those which do not.

\begin{figure} 
\epsscale{0.80}
\plotone{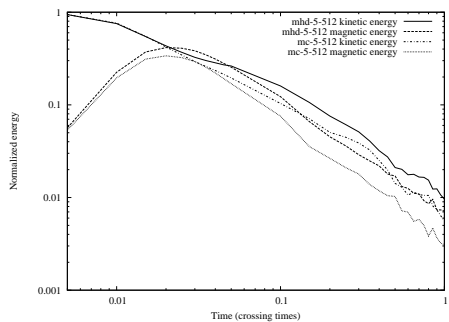}
\caption{Logscale plot of the magnetic and kinetic energies (normalized to the 
initial kinetic energy) for simulations mc-5-512 and mhd-5-512 for a
larger range of times.  See text.\label{fig:early-time-energy}}
\end{figure}

\begin{figure} 
\epsscale{0.80}
\plotone{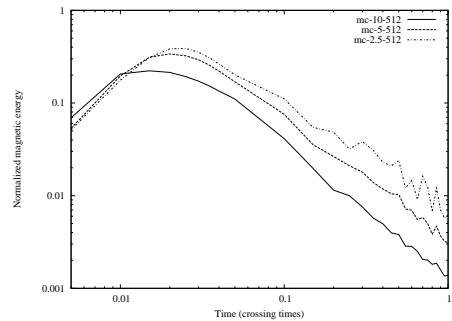}
\caption{Plot of the growth and decay of magnetic energy (normalized to the 
initial kinetic energy) for simulations mc-10-512, mc-5-512 and
mc-2.5-512.  \label{fig:early-energy-mach}}
\end{figure}

To gain a little more insight into the interplay between the magnetic and 
kinetic energy in the system we now focus on simulations mc-5-512 and 
mhd-5-512 at early times.  Figure \ref{fig:early-time-energy} contains plots 
of the magnetic and kinetic energies of these two simulations as a function  
of time.  The magnetic energy grows initially as the flow converges and 
compresses the magnetic field in regions throughout the computational 
domain.  The kinetic energy gradually decays during this time.  Once shocks 
form, equipartition between the total magnetic energy and kinetic energy is 
reached and the magnetic field begins to decay.  At this time the decay
of the kinetic energy accelerates due to the dissipative effect of the
shocks which have just formed.  This dependence on shocks forming is confirmed 
when the growth and decay of the magnetic energy is examined for simulations 
mc-2.5-512 and mc-10-512 and compared with mc-5-512 (see figure 
\ref{fig:early-energy-mach}).  Here we can see that the peak in magnetic 
energy is reached at later times as the initial rms Mach number is decreased.  
The time at which shocks begin to form is determined by the initial rms Mach 
number (giving us a characteristic velocity) and half the shortest wavelength 
in the initial velocity field.  The approximate shock formation times are 
0.0125\,$t_{\rm c}$, 0.025\,$t_{\rm c}$ and 0.05\,$t_{\rm c}$ for the 
mc-10-512, mc-5-512 and mc-2.5-512 simulations respectively.  These times 
match up reasonably well, particularly given the temporal granularity of the 
simulation data, with the peak of the fluctuating part of the magnetic
energy in these simulations.

Just after equipartition is reached the decay of the kinetic energies of 
mc-5-512 and mhd-5-512 begin to behave differently.  This is 
unsurprising as it is only when the magnetic field perturbations are 
reasonably strong that non-ideal effects can play a dynamically significant 
role.  Just after the magnetic energy peaks the kinetic energy of simulation 
mhd-5-512 goes through a short period during which it does not decay 
particularly rapidly - presumably a result of the transfer of energy from the 
compressed magnetic field back to the kinetic energy as the fluid
expands after the initial compressions in the simulation.  This does not 
happen in simulation mc-5-512 since the magnetic field does not become so 
compressed in the first place due to non-ideal effects.  This can be 
seen from the fact that the magnetic energy peaks at a lower level in this 
simulation.  In addition, energy from this field will not be transferred 
efficiently back to the kinetic energy because of the imperfect coupling 
between the magnetic and velocity fields resulting from the non-ideal terms in 
the induction equation.

The picture which emerges is the following: initially kinetic energy is 
transferred into magnetic energy through compressions.  If there are diffusive 
terms present in the induction equation then this extra magnetic energy will 
be dissipated and there will be less energy available to transfer back to the 
kinetic energy as the compressed regions expand.  This process will be 
repeated throughout the simulation as shocks form and dissipate.  Hence we 
expect a more rapid decay of both magnetic and kinetic energy when diffusive 
terms are present.  Recalling that the Hall effect does not actually diffuse 
magnetic energy, it is no surprise that simulation hall-5-512 is so similar in 
terms of energy decay to mhd-5-512.

Finally, we note that the peak of the fluctuating part of the magnetic energy
occurs at a slightly later time in the mhd-5-512 simulation than the
mc-5-512 one.  We explain this as follows.  As noted in Section
\ref{sec:rho-power}, the ambipolar diffusion appears to set a dissipation 
length scale corresponding to $k = 10$.  This is significantly larger than the 
dissipation scale in the ideal mhd simulation.  Hence energy in the magnetic 
field need not cascade to such short length scales in order to be dissipated 
in the mc-5-512 and ambi-5-512 simulations as in the mhd-5-512 and hall-5-512 
simulations.  Since the energy cascade to shorter length scales takes time 
(particularly at very early times when shocks are only beginning to form), the 
energy in the magnetic field can be removed from the former simulations at 
earlier times than from the latter simulations as it need not cascade so far.

\subsection{Power spectra}
\label{sec:power-spectra}
We now turn to power spectra of the simulations.  These tell us
about the scale of the structures formed and are useful in gaining some
insight into the effects of the non-ideal physics incorporated.  It is
important to stress that since these simulations are of decaying
turbulence, rather than driven turbulence, we do not necessarily expect
to get the often-quoted power-law dependence of power on wavenumber.
All the power spectra presented have been calculated at $t = t_{\rm c}$
when we can be fairly confident that the turbulence is fully developed
and the initial conditions have been effectively forgotten.  In
addition, the concerns noted in section \ref{subsec:resolution-study}
should be borne in mind - i.e.\ the turbulent inertial range is likely
to be of order half a decade in $k$ for our $512^3$ simulations.

We discuss the power spectra of density, magnetic field strength and
velocity in turn.

\subsubsection{Density power spectra}
\label{sec:rho-power}
%

\begin{figure} 
\epsscale{0.80}
\plotone{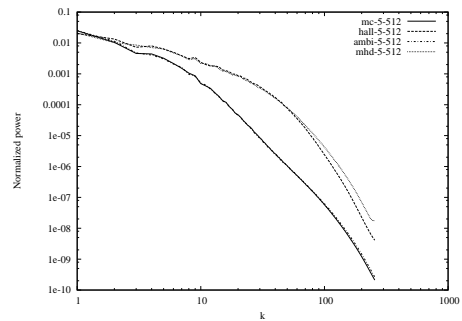}
\caption{Logscale plot of the normalized spherically integrated power 
spectrum of the density distribution for the mc-5-512, hall-5-512,
ambi-5-512 and mhd-5-512 simulations at $t = t_{\rm c}$.\label{fig:rho-power}}
\end{figure}

Figure \ref{fig:rho-power} contains plots of the spherically integrated power 
spectrum of the density for the mc-5-512, hall-5-512, ambi-5-512 and
mhd-5-512 simulations at $t=t_{\rm c}$.  It is clear that there is little 
difference between the mc-5-512 and ambi-5-512 simulations - again strongly 
indicating that the Hall effect has little influence on the behavior of the 
simulations.  The mhd-5-512 and hall-5-512 simulations are rather similar, 
although there is less power in small-scale structures in the hall-5-512 
simulation than the mhd-5-512 one.  We leave discussion of this until
section \ref{sec:b-power}.

Even at large length scales there is a significant difference in the
power spectra, indicating that dissipation on short length scales due to
the non-ideal terms in the induction equation indeed affects the behavior of 
the density distribution at large scales.  The simulations with ambipolar 
diffusion have considerably steeper slopes than those without (see 
Table \ref{table:power-slopes}) in qualitative agreement with the results of 
\cite{li08}.  That the mc-5-512 simulation has less power at shorter length 
scales than mhd-5-512 can be seen by a cursory examination of figure 
\ref{fig:mc-mhd-den} - the mc-5-512 images appear more ``blurry'' than those 
from mhd-5-512 - so this result is not a surprise.

\begin{table}
\begin{center}
\caption{The values of the exponent for the power spectra of density,
velocity and magnetic field measured at $t = t_{\rm c}$.  All fits are
over the range $5 \leq k \leq 20$ unless otherwise noted.
\label{table:power-slopes}}
\begin{tabular}{lccc}
\tableline \tableline
Simulation & Density & Velocity & Magnetic field \\ 
\tableline
mc-5-512 & 2.09\tablenotemark{a}, 4.06\tablenotemark{b} & 1.47
& 2.17\tablenotemark{a}, 4.96\tablenotemark{b} \\
ambi-5-512 & 2.04\tablenotemark{a}, 4.03\tablenotemark{b} &
1.49 & 2.21\tablenotemark{a}, 4.75\tablenotemark{b} \\
hall-5-512 & 1.41 & 1.20 & 1.65 \\
mhd-5-512 & 1.45 & 1.17 & 1.59 \\ 
mc-2.5-512 & 2.37\tablenotemark{a}, 3.32\tablenotemark{b}& 2.14 &
2.86\tablenotemark{a}, 3.80\tablenotemark{b}\\
mc-10-512 & 2.20\tablenotemark{a}, 4.20\tablenotemark{b}& 1.44 &
2.59\tablenotemark{a}, 4.70\tablenotemark{b}\\
\tableline
\end{tabular}
\tablenotetext{a}{Fitted over $4 \leq k \leq 10$}
\tablenotetext{b}{Fitted over $10 \leq k \leq 100$}
\end{center}
\end{table}

Further structure in the power spectrum for mc-5-512 is apparent which
is absent in the mhd-5-512 simulation: at $k \approx 10$ there is a break to 
a steeper slope in the simulations containing ambipolar diffusion.  For the 
mhd-5-512 simulation, at about $k \approx 35$ numerical viscosity begins to 
affect the power spectrum, as evidenced by the roll over of the spectrum from 
a power-law to a steeper slope.  The same effect does not appear in the 
mc-5-512 simulation because the physical viscosity due to the non-ideal terms 
in the induction equation dominates the numerical viscosity up to much higher 
values of $k$.  While the power spectrum for the latter simulation does change 
slope at $k \approx 10$ it then maintains a strong power law up to 
$k \approx 100$.

Hence the break at $k\approx 10$ for mc-5-512 and ambi-5-512 appears to
be a physical result, rather than a numerical one.  This suggests that,
apparently in contradiction with a result of \cite{ois06}, ambipolar 
diffusion can set a length scale in turbulence - in these simulations
that length scale is about 0.02\,pc.  We discuss this further
in section \ref{sec:b-power} but note that we should be cautious about drawing 
general inferences given that the resistivities used in these simulations are 
constant in space and time.

Ultimately, at very short length scales ($k \geq 100$ or $l \leq
0.002$\,pc) all four simulations steepen significantly.  This is expected since
at these values of $k$, which correspond to lengths of less than about 5
grid zones in these simulations, numerical viscosity will certainly dominate 
structure generation/dissipation.  

\subsubsection{Velocity power spectra}
\label{sec:vel-power}
\begin{figure} 
\epsscale{0.80}
\plotone{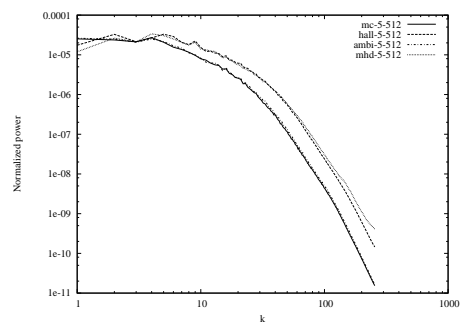}
\caption{Logscale plot of the normalized spherically integrated power 
spectrum of the velocity distribution for the mc-5-512, hall-5-512,
ambi-5-512 and mhd-5-512 simulations at $t = t_{\rm c}$.\label{fig:v-power}}
\end{figure}

We now turn to the velocity power spectra. Figure \ref{fig:v-power}
contains plots of the spherically integrated power spectrum of the
velocity for the mc-5-512, hall-5-512, ambi-5-512 and mhd-5-512 simulations at 
$t=t_{\rm c}$.  In broad terms of the differences between the simulations we 
obtain the same result as for the density power spectrum (cf figure
\ref{fig:rho-power}).  It is, however, obvious that the density and
velocity power spectra differ quite substantially in qualitative terms.

At very low $k$ the velocity spectra have a rather shallow slope with little
difference between any of the simulations up to $k \approx 4$ - the
maximum value of $k$ at which the initial velocity field was non-zero.
In the range $5 \leq k \leq 20$ both mhd-5-512 and hall-5-512 follow a
similar power law, while mc-5-512 and ambi-5-512 attain a steeper slope
in this range.  In general, for $k \geq 5$ the simulations with ambipolar 
diffusion have significantly steeper power spectra, as observed in the density 
power spectra and in qualitative agreement with the driven turbulence
results of \cite{li08}.  

There is a further apparent break in the power spectrum at around 
$k \approx 100$ in the simulations with ambipolar diffusion occurring
where numerical viscosity begins to dominate the ambipolar diffusion
(see section \ref{sec:rho-power}).  It is interesting to note that the
series of breaks in the velocity power spectra in mc-5-512
and ambi-5-512 are not mirrored in the density power spectra indicating
a level of decoupling between these two fields.

\begin{figure} 
\epsscale{0.80}
\plotone{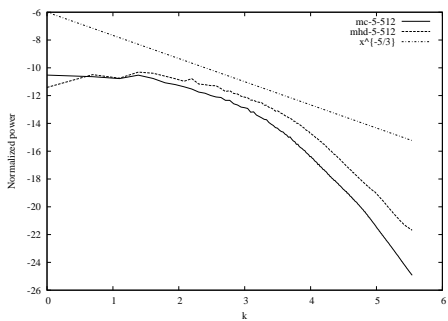}
\caption{Logscale plot of the normalized spherically integrated power 
spectrum of $\rho^{1/3}q$ for the mc-5-512 and mhd-5-512 simulations at 
$t = t_{\rm c}$.  Also shown is the $k^{-5/3}$ Kolmogorov power-law.  
\label{fig:rho-third-v-power}}
\end{figure}

\cite{krit07} noted that under certain circumstances a power spectrum of
$\rho^{1/3}q$ could follow a power law with the classical Kolmogorov
slope of -5/3 even for compressible (hydrodynamic) turbulence.  Figure
\ref{fig:rho-third-v-power} contains power spectra of this variable for the 
mc-5-512 and mhd-5-512 simulations.  It is apparent that the power
spectra for this variable for these simulations is shallower than the
-5/3 law, at least up to the value of $k$ where dissipative effects may
be important.  In the range $4 \leq k \leq 10$ the exponents are -1.289 and 
-1.14 for the mc-5-512 and mhd-5-512 simulations, respectively.  The theory 
under which these variables could be expected to have the Kolmogorov slope has 
the assumption that the system has reached a statistical steady state -
since we investigate decaying turbulence here this is unlikely to be the
case.  It might be expected that if we do not continually supply energy at low 
$k$ then the slope after a turbulent crossing time when the turbulence is
well established and has also decayed significantly would be too shallow
as the power spectra would be ``too low'' at low $k$.

\subsubsection{Magnetic field power spectra}
\label{sec:b-power}

\begin{figure} 
\epsscale{0.80}
\plotone{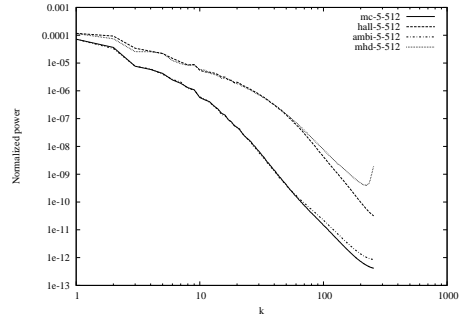}
\caption{Logscale plot of the normalized spherically integrated power 
spectrum of the magnetic field strength distribution for the mc-5-512, 
hall-5-512, ambi-5-512 and mhd-5-512 simulations at $t = t_{\rm c}$.
\label{fig:b-power}}
\end{figure}

Figure \ref{fig:b-power} contains plots of the power spectrum of the
magnetic field at time $t=t_{\rm c}$.  Once again these plots
indicate that ambipolar diffusion is the dominant non-ideal effect in molecular
cloud turbulence.  We can see that there are some differences between
each of the simulations.  The mc-5-512 and ambi-5-512 simulations are
rather similar and are rather similar to the density power spectra
(figure \ref{fig:rho-power}).  However, at high $k$ the ambi-5-512
simulation does have noticeably more power than mc-5-512.  This indicates that 
the Hall effect, present in the mc-5-512 simulation, is having some impact in 
the structure of the magnetic field at short length scales.  This effect is 
very small in the power spectra of both velocity and density and is not 
apparent in the energy decay rates either.  It would appear that when the 
ambipolar diffusion is strong in comparison to the Hall effect, as is the case 
for molecular clouds, the Hall effect may change the structure of the magnetic 
field but this change will not propagate into the rest of the fluid
variables.

The power spectra for mhd-5-512 and hall-5-512 are rather similar to
each other, again with the exception of high $k$ where the hall-5-512
simulation has significantly less power.  We explain this as follows.  While 
the Hall effect does not actually diffuse the magnetic field it does re-orient 
it.  This process of re-orientation can give rise to a topology
favoring some magnetic reconnection - particularly on the small scales at 
which the Hall effect operates.  We therefore attribute the
lower power at high $k$ in hall-5-512 to the interplay between the Hall
effect and numerical viscosity at these length scales.  Recall that
since hall-5-512 contains no ambipolar diffusion and negligible parallel
resistivity any reconnection which occurs must be almost entirely due to 
numerical viscosity.  The same argument holds for the difference between
mc-5-512 and ambi-5-512 at high $k$, but in this case the interplay is
between the Hall effect and ambipolar diffusion.  The marked turn up in power 
for $k \geq 200$ for the mhd-5-512 simulation is due to numerical effects.

Overall the plots are rather similar to the density power spectra 
(figure \ref{fig:rho-power}): the density and magnetic field distributions 
appear to be fairly well coupled and different in nature to the velocity
distribution.

\subsection{Velocity dispersion}
\label{sec:vel-disp}

It has been generally accepted that observations of line of sight velocity 
dispersion in molecular clouds exhibits a power law with the size of the 
field of view \citep[e.g.][]{lar81}.  It is worth noting, however, that
recent observational results call the so-called Larson's law into
question \citep{hey09}.  However, for completeness we feel it is
worthwhile to examine how our non-ideal simulations behave in this
regard.  In this section we examine the velocity dispersion as 
a function of characteristic length for each of the simulations.  

\begin{figure} 
\epsscale{0.80}
\plotone{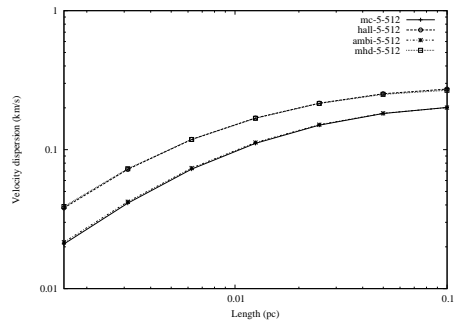}
\caption{Logscale plot of the velocity dispersion for the mc-5-512, 
hall-5-512, ambi-5-512 and mhd-5-512 simulations.  Note that the data for
simulations mhd-5-512 and hall-5-512 are almost identical.
\label{fig:larson}}
\end{figure}

Figure \ref{fig:larson} contains plots of the velocity dispersions at $t
= t_{\rm c}$ for each of the $512^3$ simulations.  Once again,
the results follow the general trend of indicating that the Hall effect 
has almost no impact with the data for mhd-5-512 and hall-5-512 being
almost indistinguishable.  At all length scales the velocity dispersion for the 
simulations containing ambipolar diffusion (mc-5-512 and ambi-5-512) is lower, 
with the difference being larger at short length scales.

From the power spectra study in section \ref{sec:power-spectra} we already know 
that the power in short scale variations of the velocity is decreased by
the presence of ambipolar diffusion.  It is therefore no surprise that
comparison of the results for the velocity dispersion shows that the
dispersion is decreased by the presence of this diffusion.

There is no obvious power law in these results, in agreement with the
simulations of driven, ideal MHD turbulence of \cite{lem09}.  As noted
by these authors, the varying strength and direction of the magnetic
field give rise to a large number of signal speeds within the domain, rather 
than the simple sound speed frequently assumed in hydrodynamic turbulence and 
hence there is no reason to expect a power law in the velocity dispersion.

\subsection{The effect of initial rms Mach number}
\label{sec:mach-effect}

Finally we turn to the issue of the initial Mach number chosen in the
simulations.  To study this we consider simulations mc-2.5-512, mc-5-512
and mc-10-512.  These correspond to initial sonic Mach numbers of 2.5, 5
and 10, respectively and Alfv\'enic Mach numbers of 0.96, 1.9 and 3.85.

\subsubsection{Energy decay}
\label{sec:mach-energy-decay}

\begin{figure} 
\epsscale{0.80}
\plotone{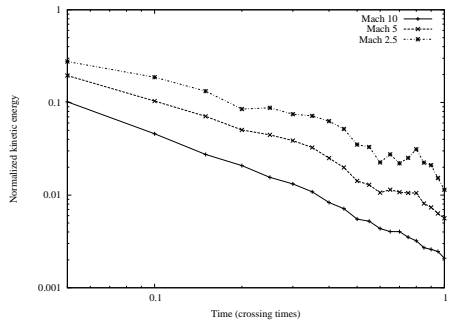}
\caption{Logscale plot of the normalized kinetic energy for the mc-10-512, 
mc-5-512 and mc-2.5-512 simulations.
\label{fig:mach-number-kin}}
\end{figure}

Figure \ref{fig:mach-number-kin} contains plots of the decay of kinetic
energy for each simulation.  The energy in the plots has been normalized
to the starting energy for each simulation.  It is clear that the higher
Mach number flows lose their energy more rapidly than the lower Mach
number ones.  This behavior is repeated in the magnetic energy and the
total energy.  Table \ref{table:decay-exp} contains the power law indices of 
decay for the kinetic, magnetic and total energy in these simulations.
The exponents for the kinetic energy decay for the mc-2.5-512, mc-5-512 and 
mc-10-512 simulations are 1.21, 1.40 and 1.42 respectively, while those
for magnetic energy are 1.29, 1.37 and 1.39 respectively.  It is
clear that, indeed, all the indices increase with initial rms Mach
number.  We explain this by noting that high Mach number flows tend to have 
strong shocks which will dissipate energy more effectively than lower Mach 
number flows.

\subsubsection{Power spectra}
\label{sec:mach-effect-power-spectra}
\begin{figure} 
\epsscale{0.80}
\plotone{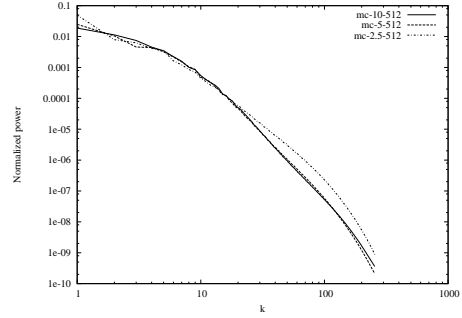}
\caption{Logscale plot of the power spectrum of the density for the
mc-10-512, mc-5-512 and mc-2.5-512 simulations.
\label{fig:mach-effect-rho-power}}
\end{figure}

\begin{figure} 
\epsscale{0.80}
\plotone{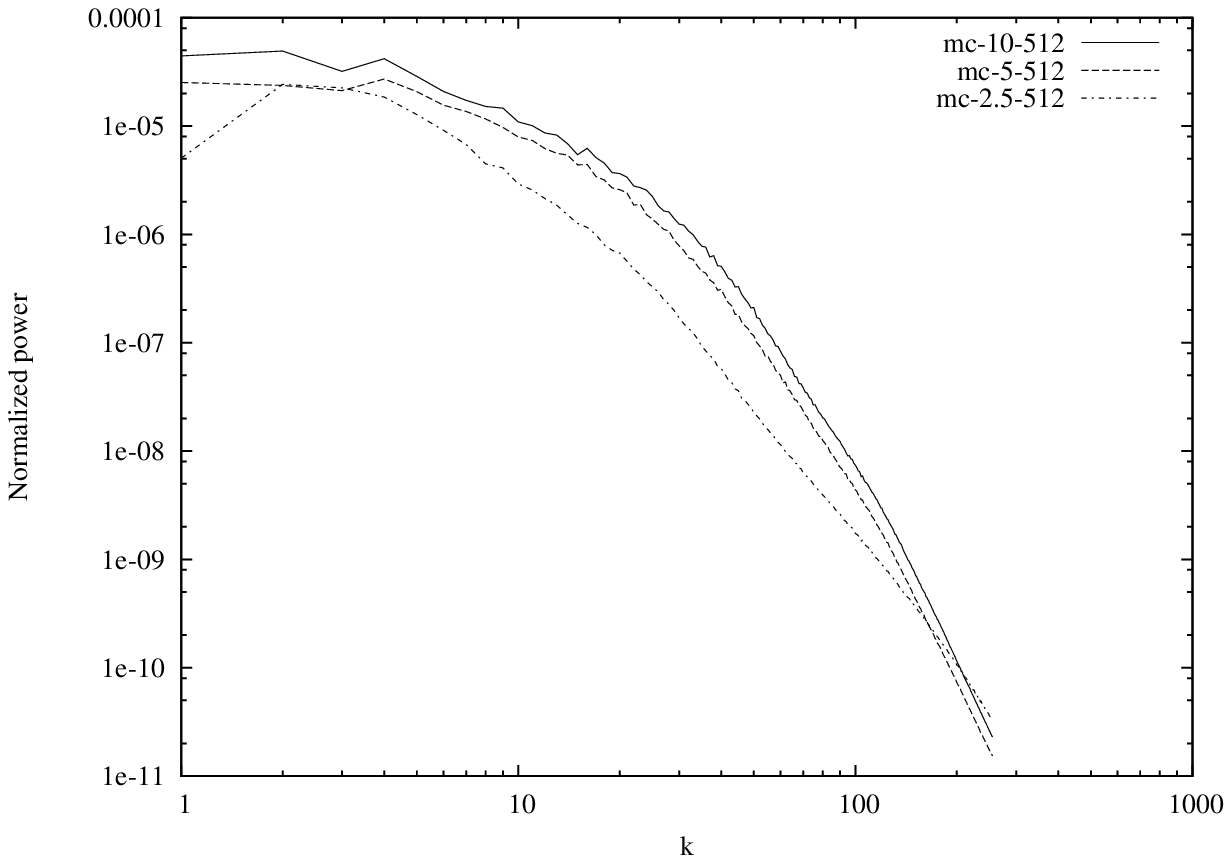}
\caption{Logscale plot of the power spectrum of the velocity for the
mc-10-512, mc-5-512 and mc-2.5-512 simulations.
\label{fig:mach-effect-v-power}}
\end{figure}

\begin{figure} 
\epsscale{0.80}
\plotone{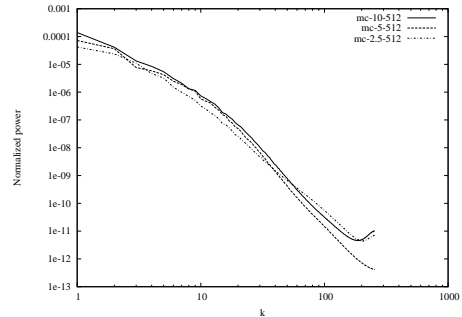}
\caption{Logscale plot of the power spectrum of the magnetic field for the
mc-10-512, mc-5-512 and mc-2.5-512 simulations.
\label{fig:mach-effect-b-power}}
\end{figure}

Figures \ref{fig:mach-effect-rho-power}, \ref{fig:mach-effect-v-power}
and \ref{fig:mach-effect-b-power} contain the power spectra for the
mc-10-512, mc-5-512 and mc-2.5-512 simulations for each of the density,
velocity and magnetic field respectively taken at $t=t_{\rm c}$.

The density power spectrum does not appear to change much between the
mc-10-512 and mc-5-512 simulations, although it is markedly shallower
for the mc-2.5-512 simulation.  It is worth noting that the mc-2.5-512
simulation is initially sub-Alfv\'enic and so is qualitatively different
in nature to the other two.  In addition, the rms sonic Mach number drops 
below 1 at $t\approx0.11$\,$t_{\rm c}$ and hence we have rather well-evolved 
subsonic, decaying turbulence in this case.

The velocity power spectrum for the mc-10-512 and mc-5-512
simulations are also remarkably similar, while the mc-2.5-512 simulation
is steeper at low $k$ and shallower for $20 \leq k \leq 100$.  Once
again, we attribute this difference in behavior to the fact that the
initial rms velocity of the mc-2.5-512 simulation is sub-Alfv\'enic.

The magnetic power spectrum follows a similar pattern in terms of
differences between the mc-10-512, mc-5-512 and mc-2.5-512 simulations
as for the other two sets of power spectra.  

\subsubsection{Velocity dispersion}

\begin{figure} 
\epsscale{0.80}
\plotone{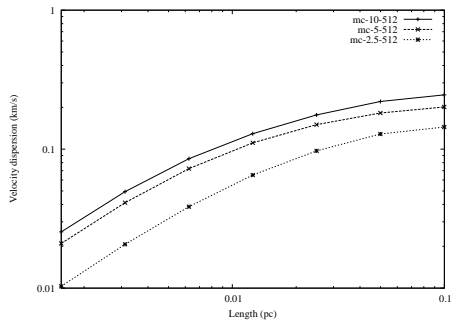}
\caption{Logscale plot of the velocity dispersion as a function of
spatial scale at $t=t_{\rm c}$ for the mc-10-512, mc-5-512 and 
mc-2.5-512 simulations.
\label{fig:mach-effect-vel-disp}}
\end{figure}

Figure \ref{fig:mach-effect-vel-disp} shows plots of the velocity
dispersion at $t=t_{\rm c}$ for the mc-10-512, mc-5-512 and
mc-2.5-512 simulations.  Again, as noted in section \ref{sec:vel-disp}, no 
overall power-law is observed.  At short length scales (less than about
0.01\,pc) the slope of the relations are all approximately the same.  Above this 
scale the slope is somewhat lower for mc-10-512 and mc-5-512 than for
mc-2.5-512.  This is to be expected since large velocity
variations will be preferentially suppressed by strong shocks during the
early evolution of the system, leading to each of these simulations having 
more similar velocity dispersion at large length scales.  Hence we
expect higher Mach number simulations to retain somewhat higher
velocity dispersions at all length scales than their lower Mach number
counterparts, but that the fractional differences in these dispersions will be 
lower at large length scales than at shorter ones.

\section{Conclusions}
\label{sec:conclusions}

We have presented the first study of turbulent decay in the presence of
non-ideal terms in the induction equation and also the first simulations
to incorporate both the Hall effect and ambipolar diffusion
simultaneously in simulations of molecular cloud turbulence.  This is
the first stage in a comprehensive study of non-ideal MHD turbulence.
The non-ideal effects are therefore included in a simplistic way in this work 
with a view to developing a complete, intuitive understanding of their impact 
on turbulence as we continue the study and gradually add in more
realistic components of these effects.  The results of this further
stage of the study are the subject of a forthcoming paper currently in
preparation.


We have used a resolution study to determine that our simulations are
well-resolved for the purposes of the results which we present.  In
particular, the turbulent energy decay is well resolved from a resolution of
about $256^3$ and the results we present here are derived from $512^3$
simulations.  

We have found that the Hall effect has little influence on kinetic energy decay 
when present at the levels found in molecular clouds.  However, ambipolar 
diffusion increases the rate of energy decay on length scales of $0.2$\,pc or 
less.  Similar results are found for the behavior of magnetic energy.  
Ambipolar diffusion increases the rate of energy decay through diffusing away 
energy while it is stored in the magnetic field.  Viewed in this way it is 
unsurprising that the Hall effect does not influence kinetic energy decay 
strongly since it does not actually diffuse magnetic energy.  The kinetic and 
magnetic energy decay is faster in higher Mach number flows as would be 
expected both from this argument and from the fact that we are simulating 
isothermal flows which will lose energy more efficiently when stronger shocks 
are present.  


The power spectra of the density for these simulations again suggest
that the Hall effect has little impact on the flows, while ambipolar
diffusion cannot be ignored.  As might be expected from a diffusion
term, the power spectrum is softer (i.e.\ steeper) when it is present.
At a resolution of $512^3$ and an assumed length-scale of $0.2$\,pc we
appear to have resolved the length at which ambipolar diffusion begins
to influence the flow.  Ambipolar diffusion does appear to impose a 
characteristic length-scale on the turbulence.  However, we must be cautious 
about this interpretation since our assumption of spatially constant 
resistivities will have some impact on this result.  

When comparing the velocity power spectra with the density power spectra
we find that there appears to be a decoupling between the two fields
with breaks in the power-laws not mirrored between the two sets of
spectra when non-ideal effects are included.

The Hall effect does have some impact on the magnetic power spectra at
high $k$.  It decreases the energy at high $k$, probably due to
re-orientation of the magnetic field at small scales in such a way as to
favor reconnection (either numerical or physical) and hence destruction of 
structure on these scales.

Calculations of the velocity dispersion as a function of length scale
show that, again, ambipolar diffusion is the dominant diffusive term and
that it has a significant impact.  It preferentially reduces the
velocity dispersion at small scales.  We do not find a power law
dependence between length scale and velocity dispersion, in agreement
with \cite{lem09}.


Comparisons of decaying turbulence with varying initial rms Mach number
show that higher Mach number flows decay more quickly than their low
Mach number counterparts.  There are also differences in the
power spectra with the Mach 2.5 flow being significantly different to
the Mach 5 and Mach 10 flows.  The Mach 2.5 flow is slightly
sub-Alfv\'enic initially and this may explain the qualitative difference
seen.  The velocity dispersions are lower at short scales for lower Mach
number flows, but are similar at large scales.  This is due to the
tendency of strong shocks to decay very quickly and hence we do not
expect even high Mach number flows to maintain high velocity dispersion
at large length scales for long.

The next step in this work is to study turbulent decay in the presence
of resistivities determined consistently from the magnetic field and the 
density of charged species throughout the computational domain.  This will 
allow us to understand precisely the impact of spatially varying resistivities 
on turbulent decay.  We can expect, for example, that the behavior of the
power spectra and velocity dispersions will be strongly affected.  In
addition, such simulations will then incorporate all the non-ideal
effects likely to be of importance in the study of molecular cloud
turbulence.

\acknowledgments{The authors would like to thank the anonymous referee for 
useful suggestions during the refereeing process.  This material is based 
upon works supported by the Science Foundation Ireland under Grant No. 
07/RFP/PHYF586.  The authors wish to acknowledge the SFI/HEA Irish Centre for 
High-End Computing (ICHEC) for the provision of computational facilities and 
support.  The work described in this paper was carried out using resources 
provided to ICHEC through the Irish National Capability Computing Initiative, 
a partnership between all the major third level research institutions and 
IBM coordinated by the Dublin Institute for Advanced Studies and supported 
by the HEA under PRTLI cycles 3 and 4 with funding from the ERDF and the NDP.}

\end{document}